# EFFECT OF UTTERANCE DURATION AND PHONETIC CONTENT ON SPEAKER IDENTIFICATION USING SECOND-ORDER STATISTICAL METHODS

*Ivan MAGRIN-CHAGNOLLEAU*°    *Jean-François BONASTRE*⋆    *Frédéric BIMBOT*°

°Télécom Paris (E.N.S.T.) – Dépt. Signal, C.N.R.S. URA 820
46, rue Barrault – F-75634 Paris Cedex 13 – FRANCE – European Union
email: ivan@sig.enst.fr and bimbot@sig.enst.fr

⋆Laboratoire d'Informatique Université d'Avignon
33, rue Louis Pasteur – F-84000 Avignon – FRANCE – European Union
email: jfb@univ-avignon.fr

## ABSTRACT

Second-order statistical methods show very good results for automatic speaker identification in controlled recording conditions [2]. These approaches are generally used on the entire speech material available. In this paper, we study the influence of the content of the test speech material on the performances of such methods, i.e. under a more analytical approach [3]. The goal is to investigate on the kind of information which is used by these methods, and where it is located in the speech signal. Liquids and glides together, vowels, and more particularly nasal vowels and nasal consonants, are found to be particularly speaker specific: test utterances of 1 second, composed in majority of acoustic material from one of these classes provide better speaker identification results than phonetically balanced test utterances, even though the training is done, in both cases, with 15 seconds of phonetically balanced speech. Nevertheless, results with other phoneme classes are never dramatically poor. These results tend to show that the speaker-dependent information captured by long-term second-order statistics is consistently common to all phonetic classes, and that the homogeneity of the test material may improve the quality of the estimates.

## 1. INTRODUCTION

In this paper, the influence of phonetic content on the performance of a speaker identification system is investigated. Second-order statistical methods are chosen because they provide very good results with a low quantity of computations and with a restricted quantity of speech material, provided recording conditions and channel distortions are controlled [2]. The goal of this work is to study how the performances of this family of approaches vary with the phonetic content of the test material.
Several experiments were previously reported on the relative speaker discriminating properties of phonemes. In particular, Eatock et al. [5] used a VQ codebook based approach and concluded that nasals and vowels provided the best performances on an English language database. Le Floch et al. [9] investigated the properties of AR-vector models and concluded that vowels, diphtongs and nasals provided the best performances, also on an English language database.
The specificity of the work reported in this paper comes from the fact that second-order statistical methods are used and that the training and test material are heterogeneous. The general experimental framework is the following: 15 seconds of training material coming from phonetically balanced sentences, are used to build a reference model for each speaker. Then, specific speech segments are selected from other phonetically balanced sentences in order to build a test pattern which is strongly biased towards a particular phoneme or phoneme class.
The database used for our experiments contains 67 cooperative speakers recorded in a slightly noisy environment. For each speaker, 50 French phonetically balanced sentences, i.e approximately 3 minutes of speech are recorded in a single session. This volume of speech allows to have a sufficient number of occurences for most phonemes.
The three second-order statistical measures used in the experiments are described in section 2.. Details on the speech database and on the signal analysis are given in section 3.. Section 4. reports preliminary experiments on utterance duration, and section 5. the experiments on the phonetic content. Finally, section 6. concludes on this work and gives some perspectives.

## 2. SPEAKER IDENTIFICATION MEASURES

The 3 speaker identification methods used in this work are inspired from statistical tests on covariance matrices [1], computed on acoustic parameters. The first two speaker similarity measures are directly derived from maximum likelihood Gaussian classifiers [7]. The third one is based on a sphericity test between covariance matrices [8]. The measures, asymmetric in their original form, are symmetrised as a weighted sum of the asymetric measure and its dual term. In previous work [2], this procedure was shown to improve performances.
Let $X$ and $Y$ denote two covariance matrices of a reference speaker and of a test speaker respectively, corresponding to the covariance of some spectral vectors computed along a sentence. Let $\bar{x}$ and $\bar{y}$ denote the means of the spectral vectors. Let $M$ and $N$ denote the number of spectral vectors used to estimate the covariance matrices and mean vectors, and $p$ the dimension of the spectral vectors. The mathematical expression of the three measures that we used in our experiments, are given in Table 1. Note that, once the covariance matrices X and Y are inverted, and that their determinant is evaluated, the computation of the measures requires very few operations.

## 3. DATABASE AND SIGNAL ANALYSIS

### 3.1. Database

The corpus is composed of read phonetically balanced sentences [4], the phonetic transcription of which can be found in the BDSON database. The sentences are prompted on a screen. Recording begins and ends automatically using a speech activity detector. Each sentence is recorded through a SHURE SM10A microphon, and digitized at a 16 kHz

$$\mu_G(X,Y) = \frac{1}{p}\left[\frac{M}{M+N}\,tr(YX^{-1}) + \frac{N}{M+N}\,tr(XY^{-1}) - \frac{M-N}{M+N}\,\log\left(\frac{detY}{detX}\right)\right]$$
$$+ \frac{1}{p}\left[(\bar{y}-\bar{x})^T\left[\frac{M}{M+N}\,X^{-1} + \frac{N}{M+N}\,Y^{-1}\right](\bar{y}-\bar{x})\right] - 1$$
$$\mu_{Gc}(X,Y) = \frac{1}{p}\left[\frac{M}{M+N}\,tr(YX^{-1}) + \frac{N}{M+N}\,tr(XY^{-1}) - \frac{M-N}{M+N}\,\log\left(\frac{detY}{detX}\right)\right] - 1$$
$$\mu_{Sc}(X,Y) = \frac{M}{M+N}\,\log\,tr(YX^{-1}) + \frac{N}{M+N}\,\log\,tr(XY^{-1}) - \frac{1}{p}\frac{M-N}{M+N}\,\log\left(\frac{detX}{detY}\right) - \log p$$

**Table 1.** *Expressions of the three symmetrized second-order statistical measures ("tr" denote the trace and "det" the determinant of a matrix).*

sampling frequency on 16 bits by an OROS AU22 board. The recording equipment was set up in the corridor of a university, i.e with a non-negligible background noise. The recordings are single-session. 67 speakers took part to the experiments, mostly students. They each recorded approximately 3 minutes of speech.

### 3.2. Signal Analysis

The speech analysis module extracts filterbank coefficients in the following way: a Winograd Fourier Transform is computed on Hamming windowed signal frames of 31.5 ms (i.e 504 samples) at a frame rate of 10 ms (160 samples). For each frame, spectral vectors of 24 Mel-Scale Triangular-Filter Bank coefficients are then calculated from the Fourier Transform power spectrum, and expressed in logarithmic scale. Covariance matrices and mean vectors are finally computed from these spectral vectors.

## 4. UTTERANCE DURATION

The first set of experiments investigates on the influence of utterance duration for second-order statistical methods. This allowed us to choose meaningful training and test durations for the second part of the work. Several durations for training and test are chosen: 15, 10, 6, 3 and 2 seconds for training; 10, 6, 3, 2 and 1 second for testing. For each speaker, all sentences are randomly concatenated together. The silences at the beginning and the end of sentences are not removed, but they generally do not exceed 0.1 second. First, a certain amount of speech is selected for training (for example 15 seconds). Then, the rest is segmented in several test portions, each portion having a predetermined duration, until 20 test portions are obtained unless the speech material is exhausted. As a result of this experimental design, the experiments are text-independent.

Percentages of correct identifications are given in Table 2. The best measure is always $\mu_G$, which confirms once again that the average of the spectral vectors is a significant source of speaker specific information, for good quality contemporaneous speech.

Even though the measures are symmetric, a clear asymmetry of the results can be observed: for example, with a training of 10 seconds and a test of 2 seconds, the percentage of correct identifications with $\mu_G$ is 95.3 %, while with a training of 2 seconds and a test of 10 seconds, it only reaches 88.4 %. In fact, if a training of 2 seconds is poorly representative of the speaker (for instance, if it contains a lot of silence), this has an effect on all the tests utterances from this very speaker, which affects considerably the overall score. If, conversely, a test of 2 seconds is of poor quality, it only causes 1 mistake, which has little impact on the global performance.

In the second part of our experiments, we chose a 15 second training duration, which is a guaranty to have a reliable training and a sufficient phonetic coverage. The test duration is chosen to 1 second for two reasons. Firstly, if a longer duration is chosen, the number of tests for a given phoneme is less significant. Secondly, if the percentage of correct identifications is too high, we felt that comparisons may lack statistical significance.

| Test Duration | | Training Duration | | | | |
|---|---|---|---|---|---|---|
| | | 15 s | 10 s | 6 s | 3 s | 2 s |
| 10 s (1095) | $\mu_G$ | **99.9** | **99.7** | **99.3** | **94.3** | **88.4** |
| | $\mu_{Gc}$ | 99.8 | 99.7 | 98.1 | 92.0 | 85.5 |
| | $\mu_{Sc}$ | 99.8 | 99.7 | 98.3 | 91.1 | 85.3 |
| 6 s (1294) | $\mu_G$ | **99.9** | **99.5** | **98.8** | **93.0** | **87.2** |
| | $\mu_{Gc}$ | 99.6 | 99.0 | 96.8 | 88.6 | 82.6 |
| | $\mu_{Sc}$ | 99.5 | 99.1 | 97.0 | 88.5 | 82.2 |
| 3 s (1340) | $\mu_G$ | **98.7** | **97.3** | **95.3** | **87.0** | **80.2** |
| | $\mu_{Gc}$ | 97.8 | 96.0 | 92.0 | 82.5 | 74.8 |
| | $\mu_{Sc}$ | 98.1 | 96.5 | 93.0 | 82.4 | 74.0 |
| 2 s (1340) | $\mu_G$ | **97.5** | **95.3** | **91.7** | **83.3** | **74.3** |
| | $\mu_{Gc}$ | 94.8 | 92.9 | 86.3 | 75.6 | 66.1 |
| | $\mu_{Sc}$ | 95.7 | 92.9 | 86.8 | 76.3 | 66.0 |
| 1 s (1340) | $\mu_G$ | **87.5** | **83.9** | **76.0** | **65.3** | **57.0** |
| | $\mu_{Gc}$ | 79.5 | 73.5 | 65.1 | 53.3 | 47.8 |
| | $\mu_{Sc}$ | 83.6 | 79.0 | 71.7 | 57.9 | 50.7 |

**Table 2.** *Speaker identification: results in percentage of correct identifications for different training and test durations. The numbers of tests for each test duration is indicated in parentheses.*

## 5. PHONETIC CONTENT

### 5.1. Segmentation

In order to study the influence of phonetic content on the performances of the second-order statistical methods, we used an automatic system for segmenting the speech material into specific phonemes or phonetic classes. This system of automatic localisation is based on a bottom-up acoustic-phonetic decoder, which is speaker independent [3], [10], [6]. For each sentence, this decoder proposes a set of weighted phonetic hypotheses. These hypotheses are then aligned with the phonetic transcription of the sentence, by a left-right alignment algorithm. In order to obtain a high localisation accuracy, the algorithm is tuned with a high level of rejection for uncertain alignments: in this experiment, 55 % of the sentences were rejected. As a consequence, the phonetic events selected by this procedure can be considered as highly reliable, and quite typical in their category. The localisation algorithm gives a small kernel for a recognized phoneme. So the phoneme segments were extended to 5 frames before and 5 frames after the kernel. Therefore, the segments are not only composed with frames of the given phoneme, as they may also include a small proportion of transitions.

### 5.2. Experimental Protocol

For this set of experiments, training material consists of all speech frames derived from 15 seconds of phonetically balanced sentences, i.e. no specific phonetic events are selected. It is in fact the exact same material as the one used in the previous experiment on the influence of utterance duration (see section 4.), with training durations of 15 seconds.

For the tests, specific phonetic classes and phonemes are selected on the rest of the speech material, by the procedure described in section 5.1.. For a particular phoneme, all the frames labeled as this given phoneme are concatenated together, and this material is divided into as many tests of 1 second as possible.

### 5.3. Description of the phoneme classes

Results for all phoneme classes and phonemes are not presented in this article: we chose to report only on those for which more than 40 tests were carried out.

A first class, refered to as *All*, contains all the phonemes. The experiments with test data from this class is however slightly different from the one in section 4., with 15 seconds for training and 1 second for testing: speech material in the class *All* is composed of acoustic material clearly identified as a phoneme by the segmentation process, and in particular, it does not contain silences, pauses, or other non-linguistic events.

The other classes are: *Vowels* (which contains oral and nasal vowels but not glides), *Oral Vowels*, *Nasal Vowels*, *Consonants* (which contains also glides), *Non-Nasal Consonants* (which contains all the consonants except the nasal consonants), *Nasal Consonants*, *Stop Consonants*, *Fricatives* and *Liquids+Glides* (which form a single class).

### 5.4. Results

Results for various phoneme classes and individual phonemes are given in Table 3 and Table 4 respectively. The percentage of correct identifications on all the tests is given, as well as the average of the correct identifications per speaker. The number of tests is also indicated.

### 5.5. Comments

Quite surprisingly, measure $\mu_G$ still performs better than $\mu_{Gc}$ and $\mu_{Sc}$: even though the mean vector within a phonetic class is expected to be strongly class dependent (and therefore not to match the training mean vector), it still keeps some consistence across phonetic classes.

The results for the class *All* outperform slightly those for the 15 s × 1 s experiment of section 4., probably because the speech material is, in the second case, more reliable.

A second observation is that the results for each phoneme class is higher than the result of the class *All*. It is also the case for most of the phonemes. This tends to show that a phonetically homogeneous test material benefits to the overall speaker identification performance, even though the training material does not share the same character.

In more detail, *Vowels* give better results than *Consonants*. *Nasal Vowels* outperform *Vowels* and *Oral Vowels*. *Non-Nasal Consonants*, and more particularly *Stop Consonants* or *Fricatives*, give lower performances than all *Consonants* together, whereas *Liquids+Glides* and *Nasal Consonants* yield higher scores. Note that the class *Liquids+Glides* gives the best results altogether, except with $\mu_{Sc}$ for which the classes *Nasal Consonants* and *Nasal Vowels* perform better. In what concerns individual phonemes, the best scores with $\mu_G$ are obtained with the phonemes /o/, /d/, /ε/, and /ã/, whereas /k/, /ʒ/, /u/, /s/ and /y/ give the poorest performance levels.

| Phonemes | All (1334) | | Vowels (1247) | |
|---|---|---|---|---|
| | I | M | I | M |
| $\mu_G$ | **90.6** | 90.5 | **97.1** | 97.0 |
| $\mu_{Gc}$ | 80.8 | 80.6 | 92.3 | 92.1 |
| $\mu_{Sc}$ | 83.5 | 83.4 | 92.0 | 91.7 |
| Phonemes | Oral Vowels (1206) | | Nasal Vowels (262) | |
| | I | M | I | M |
| $\mu_G$ | **96.0** | 95.9 | **98.1** | 98.5 |
| $\mu_{Gc}$ | 91.3 | 91.2 | 89.3 | 90.5 |
| $\mu_{Sc}$ | 90.5 | 90.5 | 93.1 | 92.9 |
| Phonemes | Consonants (1247) | | Non-Nasal Cons. (1186) | |
| | I | M | I | M |
| $\mu_G$ | **96.2** | 96.2 | **94.7** | 94.8 |
| $\mu_{Gc}$ | 91.1 | 91.0 | 89.4 | 89.3 |
| $\mu_{Sc}$ | 91.6 | 91.3 | 89.0 | 89.0 |
| Phonemes | Nasal Cons. (390) | | Stop Cons. (693) | |
| | I | M | I | M |
| $\mu_G$ | **96.9** | 97.7 | **94.2** | 95.7 |
| $\mu_{Gc}$ | 85.9 | 87.6 | 91.2 | 92.6 |
| $\mu_{Sc}$ | 95.1 | 93.9 | 91.3 | 92.7 |
| Phonemes | Fricatives (486) | | Liquids + Glides (277) | |
| | I | M | I | M |
| $\mu_G$ | **92.2** | 92.8 | **98.9** | 98.8 |
| $\mu_{Gc}$ | 83.7 | 84.8 | 92.4 | 92.3 |
| $\mu_{Sc}$ | 86.2 | 86.9 | 92.4 | 91.4 |

**Table 3.** *Speaker identification with speech material selected from specific phonetic classes. Training is composed of 15 s of phonetically balanced speech and test is composed of 1 s of phonetically biased speech. I = Global correct identification score, M = Average correct identification score over all speakers. The number of tests for each test configuration is indicated in parentheses.*

### 6. CONCLUSION

Our experiments on the effect of the phonetic content on speaker identification using second-order statistics, tend to show that, on our database, the phonetic homogeneity of the test material is usually a significant factor of improvement, even if the training material is heterogeneous. The phonetic classes that yield particularly good results are *Liquids+Glides*, *Vowels* (and more particularly *Nasal Vowels*), and *Nasal Consonants*, but results for other classes are never dramatically poor. There may therefore exist some kind of speaker-dependent *tie* between acoustic distributions across phonemes that is captured by the second-order statistical methods. This hypothesis has to be confirmed on other types of speech data, in particular noisy speech and non-contemporaneous recordings.

| Phonemes | /i/ (174) | | /e/ (105) | |
|---|---|---|---|---|
| | I | M | I | M |
| $\mu_G$ | **92.0** | 92.2 | **96.2** | 97.5 |
| $\mu_{Gc}$ | 79.9 | 81.3 | 86.7 | 88.0 |
| $\mu_{Sc}$ | 81.0 | 82.5 | 87.6 | 88.9 |
| Phonemes | /ɛ/ (258) | | /y/ (66) | |
| | I | M | I | M |
| $\mu_G$ | **97.3** | 98.2 | **87.9** | 90.6 |
| $\mu_{Gc}$ | 89.9 | 90.8 | 84.8 | 89.8 |
| $\mu_{Sc}$ | 90.7 | 93.0 | 86.4 | 92.2 |
| Phonemes | /ə/ (329) | | /a/ (378) | |
| | I | M | I | M |
| $\mu_G$ | **96.7** | 97.1 | **95.8** | 95.7 |
| $\mu_{Gc}$ | 91.8 | 91.8 | 87.8 | 86.9 |
| $\mu_{Sc}$ | 91.2 | 90.9 | 90.7 | 89.3 |
| Phonemes | /o/ (87) | | /u/ (75) | |
| | I | M | I | M |
| $\mu_G$ | **97.7** | 97.3 | **85.3** | 87.9 |
| $\mu_{Gc}$ | 87.4 | 88.3 | 73.3 | 77.4 |
| $\mu_{Sc}$ | 89.7 | 89.6 | 76.0 | 81.2 |
| Phonemes | /ã/ (108) | | /p/ (113) | |
| | I | M | I | M |
| $\mu_G$ | **97.2** | 97.4 | **91.2** | 92.9 |
| $\mu_{Gc}$ | 89.8 | 89.4 | 82.3 | 84.3 |
| $\mu_{Sc}$ | 94.4 | 92.9 | 87.6 | 88.7 |
| Phonemes | /t/ (196) | | /k/ (102) | |
| | I | M | I | M |
| $\mu_G$ | **90.8** | 92.3 | **82.4** | 86.0 |
| $\mu_{Gc}$ | 86.7 | 87.1 | 78.4 | 84.4 |
| $\mu_{Sc}$ | 89.3 | 90.4 | 78.4 | 81.0 |
| Phonemes | /d/ (111) | | /s/ (194) | |
| | I | M | I | M |
| $\mu_G$ | **97.3** | 97.5 | **87.1** | 89.0 |
| $\mu_{Gc}$ | 91.0 | 91.9 | 78.9 | 80.4 |
| $\mu_{Sc}$ | 92.8 | 93.8 | 84.0 | 84.3 |
| Phonemes | /v/ (42) | | /ʒ/ (48) | |
| | I | M | I | M |
| $\mu_G$ | **90.5** | 89.2 | **83.3** | 82.9 |
| $\mu_{Gc}$ | 76.2 | 77.0 | 77.1 | 76.0 |
| $\mu_{Sc}$ | 76.2 | 75.7 | 75.0 | 73.6 |
| Phonemes | /m/ (158) | | /n/ (149) | |
| | I | M | I | M |
| $\mu_G$ | **95.6** | 96.4 | **96.6** | 94.4 |
| $\mu_{Gc}$ | 79.7 | 80.3 | 87.2 | 88.0 |
| $\mu_{Sc}$ | 95.6 | 94.5 | 94.6 | 94.8 |

**Table 4.** *Speaker identification with speech material selected from specific phoneme realisations. Training is composed of 15 s of phonetically balanced speech and test is composed of 1 s of phonetically biased speech. I = Global correct identification score, M = Average correct identification score over all speakers. The number of tests for each test configuration is indicated in parentheses.*